# Network Analysis of Biochemical Logic for Noise Reduction and Stability: A System of Three Coupled Enzymatic AND Gates

Vladimir Privman, Mary A. Arugula, Jan Halámek, Marcos Pita and Evgeny Katz*

*Department of Chemistry and Biomolecular Science,*
*Department of Physics, and NanoBio Laboratory,*
*Clarkson University, Potsdam, NY 13699, USA*

**Abstract**

We develop an approach aimed at optimizing the parameters of a network of biochemical logic gates for reduction of the "analog" noise buildup. Experiments for three coupled enzymatic AND gates are reported, illustrating our procedure. Specifically, starch — one of the controlled network inputs — is converted to maltose by β-amylase. With the use of phosphate (another controlled input), maltose phosphorylase then produces glucose. Finally, nicotinamide adenine dinucleotide ($NAD^+$) — the third controlled input — is reduced under the action of glucose dehydrogenase to yield the optically detected signal. Network functioning is analyzed by varying selective inputs and fitting standardized few-parameters "response-surface" functions assumed for each gate. This allows a certain probe of the individual gate quality, but primarily yields information on the relative contribution of the gates to noise amplification. The derived information is then used to modify our experimental system to put it in a regime of a less noisy operation.

*Corresponding author: e-mail ekatz@clarkson.edu, phone +1-315-268-4421.

## 1. Introduction

We report a theoretical design, as well as an illustrative experimental realization for a network of three biocomputing gates, advancing an approach towards noise-reduction for robust functioning of biochemical logic based on enzymatic reactions. Biochemical information processing promises applications in medical testing/sensing/transduction, drug delivery, and implantable devices. One of the primary research challenges has been the increase of the complexity of "decision making" based on biochemical reactions, prior to interfacing with the ordinary electronics in externally connected information/signal processing/sensing devices.

Recently, we have initiated[1] a research effort to develop and adapt from device-science, analog and digital optimization techniques in order to advance towards fault-tolerant, scalable networking of biocomputing gates. As a network that processes information becomes large, and the information is processed in greater quantities and at higher levels of complexity, noise inevitably builds up and can ultimately fully degrade the useful signal which is the intended result of the computation. At this point the development of approaches to achieve what is known as fault-tolerant information processing that involves noise suppression is needed.

Biochemical computing[1-5] — in our case, based on enzymatic reactions[1,3-5] — attempts to process information with biomolecules and biological objects. However, the network-design approach used specifically for enzyme-based biocomputing has been the analog/digital information processing paradigm of ordinary electronics. Indeed, most biochemical computing studies have attempted to realize[4] and, most recently, network[5] gates that carry out Boolean digital logic functions.

We note that several chemical-computing studies have used less complex molecules.[6] However, biomolecules offer the advantage of typically being very selective in their chemical functions and therefore more usable in complex environments. In addition, they are expected to be more appropriate for interfacing with processes in living organisms, for potential applications.

Recent experimental advances in enzyme-based biocomputing have included not just realizations of several simple Boolean gates such as AND, OR, XOR, etc.,[3,4] but also *networking* of up to 3-4 concatenated gates presently.[5] The latter development has necessitated exploration[1,7] of noise suppression approaches for biochemical logic-gates networks.

As outlined in Sec. 2, fault tolerance within the analog/digital information processing paradigm is accomplished by suppression of the "analog" noise amplification by gate optimization[1] and network design; the latter is the subject of the present work. For larger networks, another, "digital" mechanism of noise amplification emerges.[7] It is suppressed by redundancy in network design and requires truly digital information processing and the appropriate network elements for filtering, rectification, etc. While this can be also applied to biomolecular computing systems,[7] the present sizes of the biochemical computing networks should allow exploring design and optimization issues related to suppression of the "analog" noise amplification.

In Sec. 2, we set up our theoretical formulation with application to a network of three coupled enzymatic AND gates. The experimental procedure is detailed in Sec. 3. The results for our initially chosen network parameters are presented and analyzed in Sec. 4, followed by discussion of the data for a system with a modification — suggested by the analysis — aimed at noise reduction.

## 2. Network Model for Biochemical Logic

In this section we introduce our modeling approach to network optimization. It proves convenient to reference a specific example, which is then experimentally realized, as described in Sec. 3-4. Therefore, let us first specify the network of three AND gates to be studied in detail, shown in Fig. 1. The network consists of three enzymes taking in two "input" chemicals each, and producing "output" as (one of) the resulting catalytic process product(s).

### 2.1. *Modular Network Representation*

Details of the enzymatic processes involved will be described later, in Sec. 3-4. At this point we note several observations pertinent to our approach. The ultimate goal of biocomputing is network scalability for stable information processing with large, complex, coupled sets of chemical reactions. Therefore, it is impractical to analyze in detail, properly represent, and optimize each of the involved information processing steps/gates, e.g., along the lines developed earlier.[1] It is impossible to separate out and study/model in detail each chemical or physical kinetic process involved in the network functioning, especially that for biochemical processes the kinetic rate parameters and even the detailed reaction pathways strongly depend on the chemical and physical conditions of the specific solution environment.

Therefore, we adopt a "modular" approach. We divide the network into representative logic-gate units, here three AND gates. This representation might not be precise but, for large networks, it is practical. Specifically, for our network the third (counting back from the output) gate is actually an artificial representation because we do not vary the activity of one of its inputs — water, which therefore is always held at its logic-1 value; see Fig. 1. Thus, this AND gate can be replaced by an "identity" function. A more realistic network representation for our coupled reactions will be presented and discussed in Sec. 4.3.

For each type of unit, in our case just one — the AND gates — we will develop a simple phenomenological representation of the response surface (to be defined shortly) in terms of a function of as few parameters as possible. The network optimization will then proceed as follows. First, we note that some variables that can be controlled to modify the "network machinery" — the most obvious being the enzyme concentrations — can be only varied within certain experimentally accessible ranges which include the initial values selected. We will experimentally probe the response of the network outputs (here just one) to variations in selective inputs. The obtained data will be analyzed for information on the phenomenological fitting parameters for the gates, and on the "noisiness" of the gates, the latter measuring their relative role in noise amplification. This information will be then utilized to adjust one or more of the controllable quantities to improve the network stability. In what follows, we detail this procedure and illustrate it on an example of our three-gate network.

## 2.2. *Single Boolean Gate Function*

Let us consider a single AND gate for definiteness: two chemicals, of concentrations $A$ and $B$, are the input (at time 0) signals for a reaction that yields the third chemical, of concentration $C$, as the output signal (the latter measured at the fixed final time, $t_{max}$, or, as common for biochemical reactions, as a slope of the signal curve over a certain time interval of approximately linear behavior). We define the Boolean 0 as zero concentrations of all the chemicals, whereas the Boolean 1 values are $A_{max}$, $B_{max}$ and $C_{max}$. We also define the "logic" variables,

$$x = A/A_{max}, \qquad y = B/B_{max}, \qquad z = C/C_{max}. \qquad (1)$$

Of primary interest in Boolean logic are the values of $x, y, z$ near 0 and 1. However, the reaction should actually be considered for all possible values of the concentrations, and described by the "response surface" function

$$z = F(x, y). \qquad (2)$$

This function depends not only on the arguments $x, y$, but also parametrically on $A_{max}$, $B_{max}$, as well as on the reaction time $t_{max}$, and other quantities, such as the reaction rates and the concentration of the enzyme, $E$, which acts as the gate "machinery." The function

$$F(x, y; A_{max}, B_{max}, t_{max}, E, ...) = C(x, y; A_{max}, B_{max}, t_{max}, E, ...) / C(1,1; A_{max}, B_{max}, t_{max}, E, ...), \qquad (3)$$

is not exactly known but only phenomenologically modeled by fitting the parameters from experimental data.[1]

Note that $C_{max} = C(x=1, y=1; A_{max}, B_{max}, t_{max}, E, ...)$, which is the denominator in Eq. 3, does not constitute an adjustable parameter. Furthermore, in fitting the function $C$, the dependence on the first four arguments in Eq. 3 is via the products $xA_{max}$, $yB_{max}$ (which are the concentrations $A, B$). Therefore, $A_{max}$, $B_{max}$ are also not adjustable parameters when fitting the reaction kinetics data. Since the gate could be used in concatenation with other gates in a logic circuit, or as a part of a sensor with fixed-range environmental inputs, etc., the values of $A_{max}$, $B_{max}$ will be set by the surroundings and therefore will also not be adjustable when we consider the logic function of the gate in the optimization step. A similar comment applies to the reaction time, which may be fixed or available for optimization only for the network as the whole. Furthermore, in enzymatic biochemistry frequently the reaction rate over an extended time interval is used as a signal, so there is no explicit time dependence. Recently, we considered a theoretical framework[1] for optimizing the set of appropriate "gate machinery" values for a single gate; here $\{E,...\}$, where the dots here and in Eq. 3 refer to kinetic parameters (e.g., reaction rate constants): The gate function adjustment in the framework of the available experimental information is designed to minimize noise buildup. Optimization of gate(s) within a network will be addressed in the following sections.

**2.3.** *Analog Noise Amplification*

There can be many sources of noise in the reaction network where our single AND gate is embedded. The input signals may not be exactly at the logic values defined for 0 and 1. The output can also have additional noise: The function $F(x,y)$ may actually have a random component in it. However, for functioning as a part of a network, the main property of concern is actually not the noise level itself but the degree of *noise amplification* in each biochemical gate function, which is due to the shape of the function $F(x,y)$ in the vicinity of the four logic argument values $x,y = (0,0), (0,1), (1,0), (1,1)$: If $F$ has large gradient values near these points, then any noise and fluctuations in the inputs $A, B$ will be amplified when manifesting as noise in the output $C$. Thus, we consider the "analog" noise in the output as resulting from small fluctuations in the input and controlled by the shape of the gate function. Noise amplification for systems presenting several gates combined together, can be prevented by having small gradient values near all four "logic" input points.

For real large-scale fault-tolerant device operation with numerous gates involved, another mode of noise build-up, termed "digital," will become important: No matter how narrow are the statistical distributions of the input signals, centered at the "logic" values, and how accurately is the function F performed experimentally, there will always be a minute probability of a really large fluctuation in the output, that will randomly yield a wrong logic (digital) value. In very large scale networks of combined gates, this "digital" error build-up mechanism becomes dominant, and additional error correction techniques based on redundancy and appropriate network design must be utilized.[7]

Our recent work[1,7] suggests that generally the introduction of network design and network elements for filtering the digital signals and for "digital" error correction may be required for networks[1] consisting more than order 10 biochemical logic processing steps. This estimate is also consistent with the one suggested by recent experiments on signal transduction by neurons.[8]

### 2.4. Phenomenological Parameterization of Gates for Stability Analysis

As pointed out in Sec. 2.1, detailed analysis of each information processing element (gate, etc.) by varying all its inputs and modeling the rate equations for all the (bio)chemical reactions, transduction steps and pathways, might be impractical as networks become large. Furthermore, this approach of "taking the network apart" may obscure possible optimization based on adjusting the *relative* activity of network elements.

Even if we fit in detail the rate constants, say $k_\alpha, k_\beta, ...$, for each gate, as considered in Sec. 2.2, then the set of the controllable parameters is $t_{\max}; E; k_\alpha, k_\beta, ...; I_\gamma, I_\delta, ...$, which actually represents a complication, as described in this paragraph. Here $I_\gamma, I_\delta, ...$ stand for input concentrations of possible additional chemicals (which are not used as our "logic" inputs) taken in by the enzymes at various stages (and possibly at various times, not just at time 0) of the network functioning. As mentioned, biochemical signals are frequently measured as rates (slopes) over a time interval, thus no specific time dependence. The rate constants can be varied primarily by changing the physical or chemical conditions, which might not always be possible in specific applications. Therefore, we can only adjust the set $E; I_\gamma, I_\delta, ...$. However, the dependence on some concentrations, especially the enzyme itself, $E$, may to a large extent cancel out in the "logic" dimensionless output, Eq. 3, because the overall gate activity is approximately linear for the enzyme concentration in many regimes of applications (with the enzymatic reactions held near steady state conditions). Thus, we are left with no gate parameters for a straightforward gate-by-gate optimization. These considerations again point to the conclusion that *relative* network optimization may be much more practical than gate-by-gate optimization.

In what follows, we consider the AND gate for definiteness, but extensions to other two-input, one-output Boolean gates are straightforward. As described in our previous work,[1] the response-surface function $z = F(x, y)$ should ideally have very small gradients near each logic point. For example for $y$ close to 1, a desirable sigmoid shape is shown in Fig. 2. However, in most situations we will have the convex shape instead, also shown in Fig. 2. This reflects the fact that for small amount of the input chemical $A\ (=xA_{\max})$, the activity of the gate will be linear in $x$. As $x$ increases, the gate activity is expected to show trend towards saturation, resulting in the convex shape. The exceptions are some of the substrates (inputs) showing self-promoter properties, typical for allosteric enzymes.

Presently, we seek the most generic, "modular" description of the common, convex response-surface shape in terms of as few parameters as possible. The formalism developed in our recent work[1] suggests that gate-function optimization in such a case involves balancing the gradients near the logic points. Not all of them can be made small at the same time. Therefore, we modify the gate-function shape to decrease some gradients, possibly increasing others, to make the largest of the four gradients near the logic points as small as possible. A more sophisticated approach,[1] not considered here, entails modeling the form of the noise (e.g., Gaussian) and calculating the spread of the noise distribution near the logic points.

For a one-variable convex function, shown in Fig. 2, the above approach thus rebalances the slopes near 0 and 1 with respect to each other: For standard rate-equation-derived functions, having smaller slope at 0 implies larger slope at 1, whereas increasing the slope at 0 will cause the slope at 1 to decrease. Therefore, we need at least one phenomenological parameter to describe this shape in the simplest way possible. A convenient fit function is

$$\text{Output}(x) = \frac{x(1+a)}{x+a}. \tag{4}$$

Here $a > 0$ can increase to $\infty$: the limit $a \to \infty$ in Eq. 4, corresponds to the linear function Output($x$) = $x$. For AND gates we will thus use the product form, Fig. 3,

$$F(x,y) = \frac{xy(1+a)(1+b)}{(x+a)(y+b)}, \tag{5}$$

which involves two adjustable positive parameters, $0 < a, b \leq \infty$.

Technically, we expect that if this proposed approximate description is accurate for a given gate, then the parameters $a(t_{\max}; E; k_\alpha, k_\beta,...; I_\gamma, I_\delta,...)$ and $b(t_{\max}; E; k_\alpha, k_\beta,...; I_\gamma, I_\delta,...)$ will be functions of the adjustable variables discussed earlier. However, without detailed rate-equation kinetic modeling, this dependence is not known, and we cannot verify that the functional form, Eq. 5, provides a good approximation for the actual $(x, y)$-dependence of the gate-function response surface. We will not attempt such a modeling in the present work. We point out, however, that the form Eq. 5, see Fig. 3, mimics the expected[1] behavior of a generic "convex" response-surface for biochemical AND gates, including the vanishing of the gradient at the logic point 00.

### 2.4. *Optimization of the Gate-Function Parameters*

The four gradient values of the fitting function in Eq. 5, are

$$\sqrt{\left(\frac{\partial F}{\partial x}\right)^2 + \left(\frac{\partial F}{\partial y}\right)^2}\Bigg|_{x=0,y=0} = 0, \qquad (6)$$

$$\sqrt{\left(\frac{\partial F}{\partial x}\right)^2 + \left(\frac{\partial F}{\partial y}\right)^2}\Bigg|_{x=1,y=0} = \frac{1+b}{b}, \qquad (7)$$

$$\sqrt{\left(\frac{\partial F}{\partial x}\right)^2 + \left(\frac{\partial F}{\partial y}\right)^2}\Bigg|_{x=0,y=1} = \frac{1+a}{a}, \qquad (8)$$

$$\sqrt{\left(\frac{\partial F}{\partial x}\right)^2 + \left(\frac{\partial F}{\partial y}\right)^2}\Bigg|_{x=1,y=1} = \frac{\sqrt{a^2 + 2a^2 b + 2a^2 b^2 + 2b^2 a + b^2}}{(1+a)(1+b)}. \qquad (9)$$

The minimum of the three values in Eq. 7-9 is obviously obtained in the symmetric case, when $a = b$, with Eq. 9 replaced by

$$\sqrt{\left(\frac{\partial F}{\partial x}\right)^2 + \left(\frac{\partial F}{\partial y}\right)^2}\Bigg|_{x=1,y=1}\Bigg|_{b=a} = \frac{\sqrt{2}a}{1+a}. \qquad (10)$$

This yields the optimal values

$$a_{\text{optimal}} = b_{\text{optimal}} = 1/(\sqrt[4]{2} - 1) \approx 5.3. \qquad (11)$$

In what follows, we will also use the following quantities,

$$A = \frac{a}{1+a}, \qquad A_{\text{optimal}} = 2^{-1/4} \approx 0.84, \qquad (12)$$

$$B = \frac{b}{1+b}, \qquad B_{\text{optimal}} = 2^{-1/4} \approx 0.84. \qquad (13)$$

We note that the actual gradient values for the optimal parameter selections, at the three logic points 01, 10, 11, are $\sqrt[4]{2} \approx 1.189$. This means that the optimized gate-functions still somewhat amplify analog noise, by approximately 19% per processing step. This property of the inherently convex gate-function response surfaces was explained earlier.[1] A somewhat different optimization, with detailed kinetic analysis, for a different gate function (with enzymes as inputs) in our previous work,[1] yielded a similar result, approximately 18% as the smallest noise amplification achievable. This gives some credence to our phenomenological parameterization, Eq. 5, which yielded a reasonable result.

**2.5. *Probing Network Response***

Let us rewrite Eq. 4 in terms of the parameter $A$ introduced in Eq. 12,

$$\text{Output}(x) = \frac{x}{(1-A)x + A}, \qquad A \in (0,1]. \tag{14}$$

Note that for $a$ varying in $0 < a \leq \infty$, the parameter $A$ varies in $0 < A \leq 1$.

We now turn our attention to the specific 3-gate network considered here, see Fig. 1. We follow the convention of numbering the gates in their sequence, counting from the output. With fixed $y_3 = 1$, the response function to the three varied inputs ($x_{1,2,3}$) can be calculated as

$$z = \frac{x_1 x_2 x_3 (1+a_1)(1+a_2)(1+a_3)(1+b_1)(1+b_2)}{(x_1+a_1)[x_2 x_3(1+a_2)(1+b_2)(1+a_3) + b_1 x_3(1+a_3)(x_2+a_2) + b_1 b_2(x_2+a_2)(x_3+a_3)]}, \tag{15}$$

where Eq. 5 was used for each gate, with the subscripts indicating the inputs according to the gates. Suppose that we set the inputs $x_{2,3} = 1$, and measure the function $z(x_1)$. In our experiments, described in Sec. 3-4, we varied the concentration of nicotinamide adenine dinucleotide (NAD$^+$) in the range corresponding to $x_1 \in [0,1]$, though in principle one could even probe this function for arguments larger than 1. Then the resulting data should be fitted to the functional form obtained from Eq. 15, where for brevity we omit all the fixed arguments,

$$z(x_1) = \frac{x_1(1+a_1)}{(x_1+a_1)} = \frac{x_1}{(1-A_1)x_1 + A_1}. \tag{16}$$

For variation of the output as a function of the other two inputs, phosphate (gate 2) and starch (gate 3), we get

$$z(x_2) = \frac{x_2(1+a_2)(1+b_1)}{x_2(1+a_2)+b_1(x_2+a_2)} = \frac{x_2}{(1-A_2B_1)x_2 + A_2B_1}, \quad (17)$$

$$z(x_3) = \frac{x_3(1+a_3)(1+b_1)(1+b_2)}{x_3(1+b_1+b_2)(1+a_3)+b_1b_2(x_3+a_3)} = \frac{x_3}{(1-A_3B_2B_1)x_3 + A_3B_2B_1}. \quad (18)$$

Interestingly, in terms of the variables $A$ and $B$, see Eq. 12, 13, not only the input-$x_1$ dependence, but also the input-$x_{2,3}$ dependences become single-parameter fits of the type of Eq. 14.

Variation of $x_1$ provides information on one of the phenomenological parameters, $a_1$, of gate 1. Variation of $x_{2,3}$ does not actually lead to a more complicated several-parameter data fit, even though the signal being varied, goes through more than one gate before affecting the output. Instead, we get information on a combination of parameters from more than one gate parameterization. The rightmost forms of Eq. 16-18 will be used to fit our experimental data.

## 3. Experimental

### 3.1. *Chemicals and Reagents*

The following biocatalysts and other chemicals were purchased from Sigma-Aldrich and used as supplied: β-amylase (to be denoted βAm) from sweet potato, Type-1 (EC 3.2.1.2); maltose phosphorylase (to be denoted MPh) from *Enterococcus* recombinant, expressed in *E. coli* (EC 2.4.1.8); glucose dehydrogenase (to be denoted GDH) from *Pseudomonas Sps.* (EC 1.1.1.47); β-nicotinamide adenine dinucleotide sodium salt from yeast (NAD$^+$); 4-(2-hydroxyethyl)-1-piperazineethanesulfonic acid (HEPES) 99.5%; starch (wheat purified) heat hydrolyzed. Sodium phosphate (dibasic anhydrous) was purchased from Fisher Chemicals. Ultrapure water (18 MΩ cm$^{-1}$) from NANOpure Diamond (Barnstead) source was used in all of the experiments.

## *3.2. Biochemical Logic Gates*

All the logic circuits have been designed according to Fig. 1, where [NAD$^+$]/[NAD$^+$]$_{max}$ is $x_1$, [phosphate]/[phosphate]$_{max}$ is $x_2$, [starch]/[starch]$_{max}$ is $x_3$, [β-glucose]/[β-glucose]$_{max}$ is $y_1$, [α-maltose]/[α-maltose]$_{max}$ is $y_2$, and the "activity" of water is proportional to $y_3$ (this input is unchanged in the present experiment). The output $z$ is related to [NADH]; see Sec. 3.3. GDH is $E_1$ and constitutes the machinery of the final AND gate, MPh is $E_2$ — the intermediate AND gate, and βAm is $E_3$ — the initial AND gate; cf. Fig. 1.

The enzymatic reactions can be summarized by

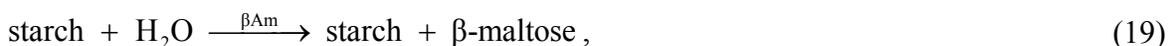
$$\text{starch} + \text{H}_2\text{O} \xrightarrow{\beta\text{Am}} \text{starch} + \beta\text{-maltose}, \tag{19}$$

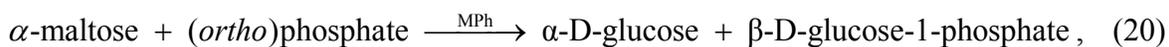
$$\alpha\text{-maltose} + (ortho)\text{phosphate} \xrightarrow{\text{MPh}} \alpha\text{-D-glucose} + \beta\text{-D-glucose-1-phosphate}, \tag{20}$$

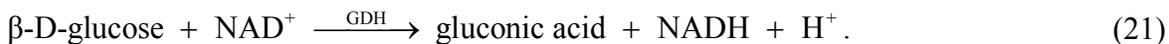
$$\beta\text{-D-glucose} + \text{NAD}^+ \xrightarrow{\text{GDH}} \text{gluconic acid} + \text{NADH} + \text{H}^+. \tag{21}$$

The kinetics of the network of three AND gates thus involves two slow interconversion steps: βAm produces[9] β-maltose, which naturally converts[9,10,11] to α-maltose in aqueous environment; MPh produces[12] α-glucose, which naturally converts[13] to β-glucose. These interconversion processes, involving equilibration between the α and β anomeric forms for each of the two saccharides, have times scales[14,15] of order 15 to 40 minutes (dependent on the process and temperature), but were observed[10] to proceed faster, on time scales of order 10 to 12 min at 37°C, in systems analogous to ours, with one of the anomeric forms constantly removed from the environment by an enzymatic reaction. To speed up the dynamics, our experiments were carried out at an elevated temperature, (50.0 ± 0.1)°C; this also makes βAm work faster.[16,17] The duration of each of our experiments was $t_{max}$ = 1000 s. (For this time scale, there was virtually no signal at room temperature. We also checked that there was no spontaneous generation of NADH from NAD$^+$ at 50°C.)

The initial (not optimized) choice of the parameter values was as follows. The experiment was carried out in 1 mL of 0.05 M HEPES buffered, pH 7.0, aqueous solution, containing the enzymes: βAm, 2 units; MPh, 2 units; GDH, 2 units. The values considered as logic-1, were $(A_{max})_1$ = 10 mM starch, $(A_{max})_2$ = 10 mM phosphate, and $(A_{max})_3 = 1$ mM NAD$^+$. The optimized system was the same, but with the amount of GDH reduced to 0.18 units, as explained in Sec. 4.

### 3.3. *Measurements*

The signal was measured by monitoring the optical absorbance, ΔA, of NADH at $\lambda = 340$ nm.[18] The biocatalytic reactions described in Sec. 3.2, were performed in a quartz cuvette, and the absorbance measurements were carried out using a UV-2401PC/2501PC UV-visible spectrophotometer (Shimadzu, Tokyo, Japan) with a thermostated cuvette chamber. Two definitions of the "output" were considered. The first one involved the *slope* (rate), $\Delta\dot{A}$, of the ΔA(*t*) curve in an approximately linear, large-time regime which was typically reached after 600 to 700 s, depending on the experiment. The second definition utilized the *value* of the absorbance of NADH, $\Delta A_{NADH}(t_{max})$, at the final time, 1000 s. Thus, we have two output definitions,

$$z_{slope} = (\Delta\dot{A})/(\Delta\dot{A})_{max}, \qquad (22)$$

$$z_{value} = [\Delta A_{NADH}(t_{max})]/[\Delta A_{NADH}(t_{max})]_{max}. \qquad (23)$$

Examples of the measured absorbance are given in Fig. 4. For evaluation of $z_{value}$, we had to devise an approach to subtract the background due to light scattering by the suspended starch. This background is a large contribution at short times, see Fig. 4, but it decays at large times. For the initial (non-optimized) system, the large-time background was rather small as compared to the signal, and we subtracted it by using the value with no input (which ranged from 0% to no more than 5% of the maximal-input signal) as the reference to set the zero of the signal. The optimized system (see Sec. 4) involved much smaller signals. Therefore, the background was subtracted by estimating the decay by using a simplified assumption that all the processes involved in the dissolution of the suspended starch due to consumption of the already dissolved starch by βAm, follow first-order kinetics. This suggests an exponential fit of the behavior of ΔA(*t*) for times for which the absorbance is dominated by the starch, typically taken as an interval from 5-10 s to 200-250 s. This fitted exponential function was then subtracted from the larger-time ΔA(*t*) values to yield $\Delta A_{NADH}(t_{max})$; see Fig. 4.

Finally, we comment that some of our measurements for the initial (not optimized) parameter choices, for larger outputs, reached values of absorbance which were out of the linear range of the Lambert-Beer law. In such cases, the "slope" signal was estimated from a smaller range of time values, selected to keep the optical density below 1.2. The "value" signal was still estimated at 1000 s, because the linearity is not required for a definition of logic-1, as long as we consistently use Eq. 23.

## 4. Results and Discussion

### 4.1. *Results for the Initial Gate Parameters*

Our first set of data was collected with the experimentally convenient, but otherwise initially randomly selected values for the adjustable "gate machinery" parameters, which we will limit to the initial enzyme concentrations, $E_{1,2,3}$, for definiteness, though, as described in Sec. 2, we could also try to vary chemical or physical conditions to attempt to influence the rates of various reactions.

The collected data were rescaled into the logic variable ranges between 0 and 1, and fitted according to Eq. 16-18. The results are summarized in Table 1, whereas the actual data and the fitted curves are shown in Fig. 5 for the slope-based signal definition, and in Fig. 6 for the fixed-time signal definition. The signal values at the maximal inputs, used to rescale the signals to the range between 0 and 1 (to convert the actual signals to the values of *z*), see Eqs. 22, 23, were $(\Delta \dot{A})_{max} = (3.8 \pm 0.2) \times 10^{-3}$ s$^{-1}$, $[\Delta A_{NADH}(t_{max})]_{max}$ = 2.9 ± 0.3, where the error bars roughly represent the actual ranges of values due to variation between the three different-input-control experiments.

For the present set of experiments, the input concentrations were not evenly spaced (see Fig. 5 and 6). The reason for this was that we wanted to carefully scan the low analog-input regime where the noise in the output signal was relatively more significant, and where the slope of the output was important for our data fits. Specifically, the concentrations ranged from 0.01 mM to 10 mM in the case of starch and phosphate, and from 0.001 mM to 1 mM for NAD$^+$.

We note that since not all the inputs of all the gates are varied to probe the response of the final output, we do not get all the 6 phenomenological fit parameters. We only get one parameter and two additional combinations of parameters. Thus, we can only draw a limited set of conclusions regarding the network noisiness. For the following discussion, the data were recast, in Table 2, in terms of the geometric means of the parameters that are known only as combinations (as products). Furthermore, since in the optimal-value case the gradients at the non-00 logic-points, discussed at in Sec. 2, are actually $1/A$ (or $1/B$) and $\sqrt{2}A$ (or $\sqrt{2}B$), cf. Eq. 7-10, we took the maxima of these quantities, derived from the geometric means, to compare to the optimal (the smallest possible) value of the gradients, $\sqrt[4]{2} \approx 1.19$.

The following semi-quantitative conclusions follow from considering the data summarized in Tables 1, 2. We note that the slope-based and the value-based signal definitions gives qualitatively similar results. Gate 3 seems to be the least noisy, whereas the larger "noise amplification measure" values that involve the other two gates should probably be attributed to Gate 1 which contributes to both measures and which is thus the primary candidate for parameter modification. In fact, the maximal values in Table 2 were *all* realized with the $1/A$ or $1/B$ type value combinations, rather than the combinations involving $\sqrt{2}A$ or $\sqrt{2}B$ type expressions. This suggests that the gradients are generally larger at logic 01 points and 10 points, as compared to logic 11 points. This means that one way to decrease noise amplification in our network is to shift the gradients from lower to higher input concentrations, i.e., work less close to saturation, i.e., slow (some of) the reactions down. Since gate 1 was already identified as the candidate for adjustment, we formulate the following theoretical "optimization recommendation:" decrease the (initial) amount of the enzyme GDH that constitutes the "machinery" of gate 1. Since our discussion in Sec. 2.4 suggested that the dependence of the logic (rescaled to the range from 0 to 1) output on the enzyme concentrations might be rather weak, we expect that a significant reduction, about on order of magnitude (by a factor comparable to 10) in the value of $E_1$, will be required.

### 4.2. *Modified Experiment with Network Optimization*

A new set of data was measured, with the reduced concentration of GDH, $E_1 = 0.18$ units. This value was selected by a preliminary set of experiments, not detailed here, carried out with the maximal values of all the inputs. The finally selected reduced values of $E_1$ was that for which there was a clear quantitative reduction in the signal intensity. Another modification was that here the data were taken with more even steps of the inputs being varied, covering in nearly equal steps the ranges from 0.5 mM to 10 mM in the case of starch and phosphate, and from 0.05 mM to 1 mM for NAD$^+$. The reason for this was that the signal being smaller in this case, the overall relative noise in the data (mostly due to the starch background) was larger even for large inputs.

The actual maximal-input signal intensities were $(\Delta\dot{A})_{max} = (1.1 \pm 0.3)\times 10^{-3}$ s$^{-1}$, $[\Delta A_{NADH}(t_{max})]_{max} = 0.6 \pm 0.2$, where, as before, the error bars roughly represent the actual ranges of values due to variation between the three different-input-control experiments. The variation here is more significant than in the large-signal regime, as is the level of noise in the data, which are analyzed in Fig. 7, 8, with the fitting results given in Table 3, and further examined in Table 4. However, as we already emphasized, we are aiming at identifying the regime of *reduced noise amplification*, even if the actual relative noise in the output signal for this particular parameter range is larger (which should be attributed to the starch background in the optical signal readout more than to the actual enzymatic network functioning). From this point of view, the results are quite promising. First, if we consistently use our simple phenomenological data fitting functions, which derive from the "convex" Eq. 5, then the noise-amplification measures (see Table 4 vs. Table 2) are consistently lower for the modified network as compared to the original one.

However, the parameter modification made, apparently took the system out of the "convex" regime where the main shape feature probed by our fit was the balance between the gradients at 01 and 10 gate-inputs vs. the 11 input. Having removed the main noise-amplifying shape, we see new features of the data, which were not apparent for the initial data set considered in Sec. 4.1. In fact, for $x_{1,2}$ as inputs (see Table 3), the assumed data fit of the type of Eq. 14, for all but one of the data sets works best with the fit-parameter value at 1 (at the edge of the expected interval of values that derive from the assumed convex gate-response-surface shapes).

Thus, the data are obviously suggestive of additional trends not captured by the simple straight-line fits in these cases. Detailed exploration of the individual gate properties, manifesting themselves in these emerging trends in the data, seen in some of the panels in Fig. 7, 8, is outside the scope of the present work because such a study would take us away from the "modular" network modeling approach, into researching the details of the actual biochemical reaction kinetics at various stages of our coupled-reaction system. A qualitative discussion along the lines of a more complicated network representation of the processes involved, at the expense of losing the modularity of the model description, is mentioned below.

Finally, we note that all but one of the "quality measure" values in Table 4 were realized as the maximal value involving combinations of the type $\sqrt{2}A$ or $\sqrt{2}B$, rather than the $1/A$ or $1/B$ type combinations (as in Table 2, see Sec. 4.1). This suggests that by taking $E_1 = 0.18$ units as an optimized-network value, we have indeed pushed the network gates into the regime away from saturation, with larger gradients at logic-11 allowing for reduction in the gradients at logic-01 and 10.

**4.3.** *Alternative Network Formulation*

Network type arguments could be advanced to qualitatively address some of the observed properties, but, as just mentioned, the network will not be "modular" and thus will not be accessible to a simple analysis of the type carried out for our 3-AND-gate representation. The modified network is presented in Fig. 9. We do not attempt its detailed analysis.

What was the "AND gate 3," is now replaced by the identity (transduction) function, I. Another I-function represents the transduction of the chemical signal to optical signal at the output stage. Note that much of the noise observed in the data (Fig. 5-8) can be attributed to this step, due to interference from starch. The functioning of the actual enzymatic network is likely much less noisy. However, as pointed out, it is not the *level* of noise *per se*, but primarily the degree of noise *amplification* what matters for stability of networks for information processing.

Two time-delayed-identity functions, denoted D, are present in the network in Fig. 9, to represent the conversion between the two anomeric forms for glucose and for maltose. These network elements act to slow down the kinetics. The presence of these time-delay steps, sets the rather large (~ 1000 s) time scales of the dynamics of the network. As a result, better resolution of the network kinetics is possible, which in principle allows for better balancing of various parameters and for avoidance of rate mismatches between the functions of different enzymes.

**4.4.** *Conclusion*

In this work we explored a modular approach that allows analysis of performance of an enzymatic network for biocomputing and other applications of biomolecule-based information processing. We proposed a methodology that, by avoiding detailed kinetic modeling for each enzyme, allows for selected probes of a complex network to be used to identify and try to adjust parameters of those gates that contribute the most to noise amplification. The adjustment in *relative activity* of the involved enzymes is a promising approach because it does not require detailed knowledge of the dependence of the modular kinetics fit-parameters on the physical and (bio)chemical environment in which the network is functioning. Our experimental study of a network of three coupled AND gates, offered an illustration of the developed theoretical ideas. Future work will be focused on studies of larger systems, those of interest in applications, as well as consideration of other network elements required to extend the analog/digital paradigm of scalable information processing from modern electronics to biochemical and hybrid information processing.

**4.5.** *Acknowledgements*

We thank O. Gromenko for useful input and discussions during the initial stages of this project. We gratefully acknowledge support of our research programs by the National Science Foundation under grants CCF-0726698 and DMR-0706209, and by the Semiconductor Research Corporation under award 2008-RJ-1839G.

# References

1. Privman, V.; Strack, G.; Solenov, D.; Pita, M.; Katz, E. *J. Phys. Chem. B* **2008**, *112*, 11777-11784.

2. a) Shao, X.G.; Jiang, H.Y.; Cai, W.S. *Prog. Chem.* **2002**, *14*, 37-46. b) Saghatelian, A.; Volcker, N.H.; Guckian, K.M.; Lin, V.S.Y.; Ghadiri, M.R. *J. Am. Chem. Soc.* **2003**, *125*, 346-347. c) Ashkenasy, G.; Ghadiri, M.R. *J. Am. Chem. Soc.* **2004**, *126*, 11140-11141. d) Stojanovic, M.N.; Stefanovic, D.; LaBean, T.; Yan, H. In: *Bioelectronics: From Theory to Applications*, Willner, I.; Katz, E. (Eds.): Wiley-VCH, Weinheim, **2005**, pp. 427-455.

3. Strack, G.; Pita, M.; Ornatska, M.; Katz, E. *ChemBioChem* **2008**, *9*, 1260-1266.

4. a) Baron, R.; Lioubashevski, O.; Katz, E.; Niazov, T.; Willner, I. *Org. Biomol. Chem.* **2006**, *4*, 989-991. b) Baron, R.; Lioubashevski, O.; Katz, E.; Niazov, T.; Willner, I. *J. Phys. Chem. A* **2006**, *110*, 8548-8553. c) Baron, R.; Lioubashevski, O.; Katz, E.; Niazov, T.; Willner, I. *Angew. Chem. Int. Ed.* **2006**, *45*, 1572-1576.

5. a) Niazov, T.; Baron, R.; Katz, E.; Lioubashevski, O.; Willner, I. *Proc. Natl. Acad. USA.* **2006**, *103*, 17160-17163. b) Strack, G.; Ornatska, M.; Pita, M.; Katz, E. *J. Am. Chem. Soc.* **2008**, *130*, 4234-4235. c) Privman, M; Tam, T. K.; Pita, M.; Katz, E. *J. Am. Chem. Soc.* In press.

6. a) De Silva, A.P. ; Uchiyama, S. *Nature Nanotechnol.* **2007**, *2*, 399-410. b) Credi, A. *Angew. Chem. Int. Ed.* **2007**, *46*, 5472-5475. c) Pischel, U. *Angew. Chem. Int. Ed.* **2007**, *46*, 4026-4040. d) De Silva, P.A.; Gunaratne, N.H.Q.; McCoy, C.P. *Nature* **1993**, *364*, 42-44.

7. Fedichkin, L.; Katz, E.; Privman, V. *J. Comput. Theor. Nanoscience* **2008**, *5*, 36-43.

8. a) Feinerman, O.; Rotem, A.; Moses, E. *Nature Physics* **2008**, *4*, 967-973. b) Feinerman, O.; Moses, E. *J. Neurosci.* **2005**, *26*, 4526-4534.

9. Weise, S.E.; Kim, K.S.; Stewart, R.P.; Sharkey, T.D. *Plant Physiology* **2005**, *137*, 756-761.

10. Shirokane, Y.; Ichikawa, K.; Suzuki, M. *Carbohydr. Res.* **2000**, *329*, 699-702.

11. Tsumuraya, Y.; Brewer, C.F.; Hehre, E.J. *Arch. Biochem. Biophys.* **1990**, *281*, 58-65.

12. Hüwel, S.; Haalck, L.; Conrath, N.; Spener, F. *Enzyme Microb. Technol.* **1997**, *21*, 413-420.
13. Nelson, D.L.; Cox M.M. *Lehninger Principles of Biochemistry*, 4$^{th}$ ed.: W.H. Freeman and Company, New York, **2005**, p. 242.
14. Shirokane, Y.; Suzuki, M. *FEBS Lett.* **1995**, *367*, 177-179.
15. Pagnotta, M.; Pooley, C.L.F.; Gurland, B.; Choi, M. *J. Phys. Organic Chem.* **1993**, *6*, 407-411.
16. Germain, P.; Crichton, R.R. *J. Chem. Technol. Biotechnol.* **1988**, *41*, 297-315.
17. Ohba, R.; Shibata, T.; Ueda, S. *J. Ferment. Technol.* **1979**, *57*, 146-150.
18. *Methods of Enzymatic Analysis*, Bergmeyer, H.U. (Ed.), Vol. 4, 2$^{nd}$ ed.: Academic Press, New York, **1974**, pp. 2066-2072.

**Table 1.** Results of the experimental data fits for the initially selected parameter set.

| Input Varied | Fitted Parameter | For Slope Signal | For Fixed-Time Signal |
|---|---|---|---|
| $x_1$ | $A_1$ | $0.66 \pm 0.02$ | $0.52 \pm 0.02$ |
| $x_2$ | $A_2 B_1$ | $0.24 \pm 0.02$ | $0.41 \pm 0.02$ |
| $x_3$ | $A_3 B_2 B_1$ | $0.50 \pm 0.05$ | $0.56 \pm 0.05$ |

**Table 2.** Gate-function "quality measures" (error ranges suppressed, cf. Table 1), derived from the data fits for the initially selected parameter set.

| Input Varied | Quality Measure | Slope Signal[*] | Fixed-Time Signal[*] | Optimal-Conditions Value: $\sqrt[4]{2}$ | Gates Involved |
|---|---|---|---|---|---|
| $x_1$ | $\max[\sqrt{2} A_1 , 1/A_1]$ | 1.52 | 1.92 | 1.19 | 1 |
| $x_2$ | $\max[\sqrt{2}\sqrt{A_2 B_1} , 1/\sqrt{A_2 B_1}]$ | 2.04 | 1.56 | 1.19 | 1, 2 |
| $x_3$ | $\max[\sqrt{2}\sqrt[3]{A_3 B_2 B_1} , 1/\sqrt[3]{A_3 B_2 B_1}]$ | 1.26 | 1.21 | 1.19 | 1, 2, 3 |

[*]All the maximal values were realized as inverses, rather than as $\sqrt{2} \times$ (see text).

**Table 3.** Results of the experimental data fits for the modified parameter set.

| Input Varied | Fitted Parameter | For Slope Signal | For Fixed-Time Signal |
|---|---|---|---|
| $x_1$ | $A_1$ | $0.87 \pm 0.06$ | $1^*$ |
| $x_2$ | $A_2 B_1$ | $1^*$ | $1^*$ |
| $x_3$ | $A_3 B_2 B_1$ | $0.57 \pm 0.10$ | $0.68 \pm 0.08$ |

$^*$See text for details.

**Table 4.** Gate-function "quality measures" (error ranges suppressed, cf. Table 3), derived from the data fits for the modified parameter set.

| Input Varied | Quality Measure | Slope Signal | Fixed-Time Signal | Optimal-Conditions Value: $\sqrt[4]{2}$ | Gates Involved |
|---|---|---|---|---|---|
| $x_1$ | $\max[\sqrt{2} A_1, 1/A_1]$ | 1.23 | 1.41 | 1.19 | 1 |
| $x_2$ | $\max[\sqrt{2}\sqrt{A_2 B_1}, 1/\sqrt{A_2 B_1}]$ | 1.41 | 1.41 | 1.19 | 1, 2 |
| $x_3$ | $\max[\sqrt{2}\sqrt[3]{A_3 B_2 B_1}, 1/\sqrt[3]{A_3 B_2 B_1}]$ | $1.21^*$ | 1.24 | 1.19 | 1, 2, 3 |

$^*$This value was the only one realized as the inverse; all the other values were $\sqrt{2} \times$ (see text).

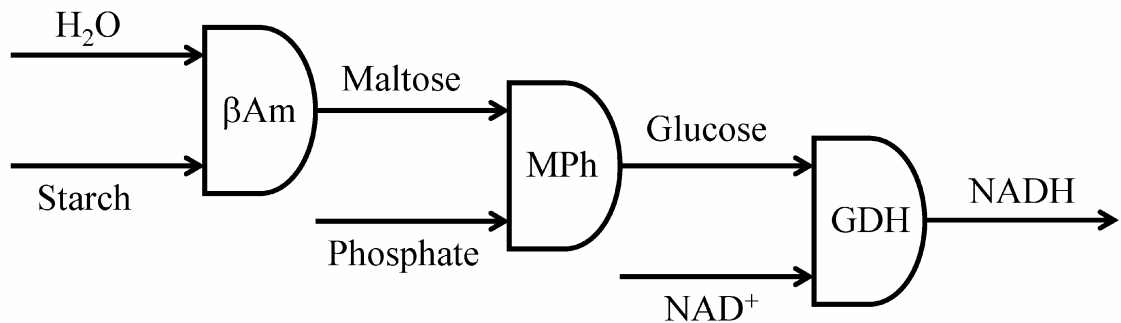

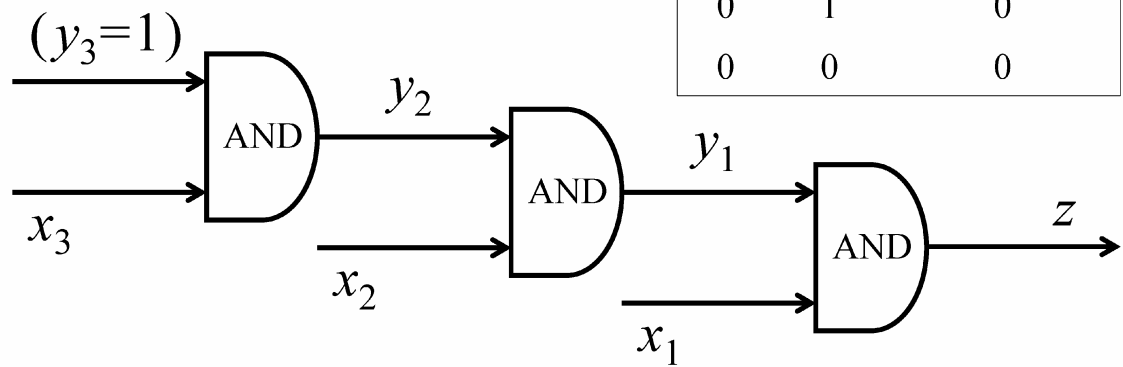

**Figure 1**. Schematic representation of the enzymatic network. Upper panel: the reagents involved as inputs/outputs, and the three enzymes used as the "gate machinery," each carrying out an AND Boolean gate. The inset spells out some of the abbreviations (with additional specifications provided in the text). Lower panel: the network of three AND gates with logic-variable inputs and outputs identified. The inset gives the "truth table" for an AND gate.

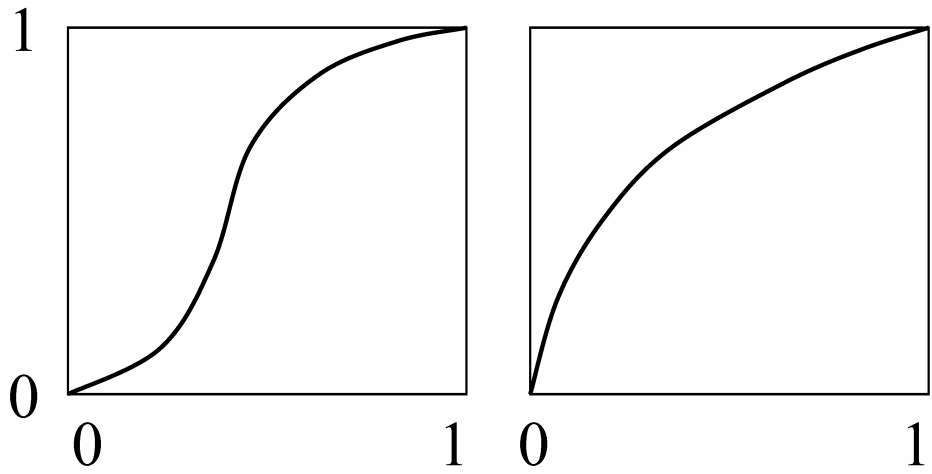

**Figure 2**. Schematic representation of cross-sections of the response surface. Left: sigmoid shape. Right: convex shape.

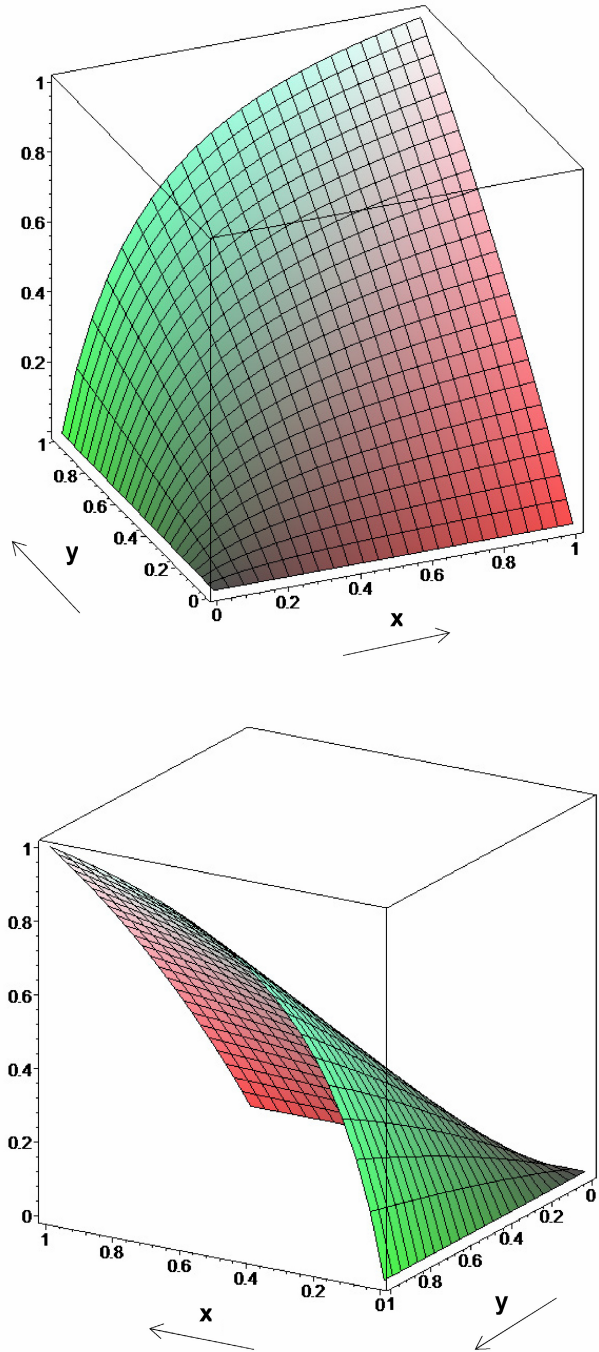

**Figure 3**. The fitting function defined in Eq. 5, shown for randomly selected representative values $a = 1/4$, $b = 4$. The two views (of the same function) illustrate the balance of the slopes of the cross-sections at output values 0 vs. 1, when one of the inputs is near 1, as well as the vanishing gradient when both inputs are 0.

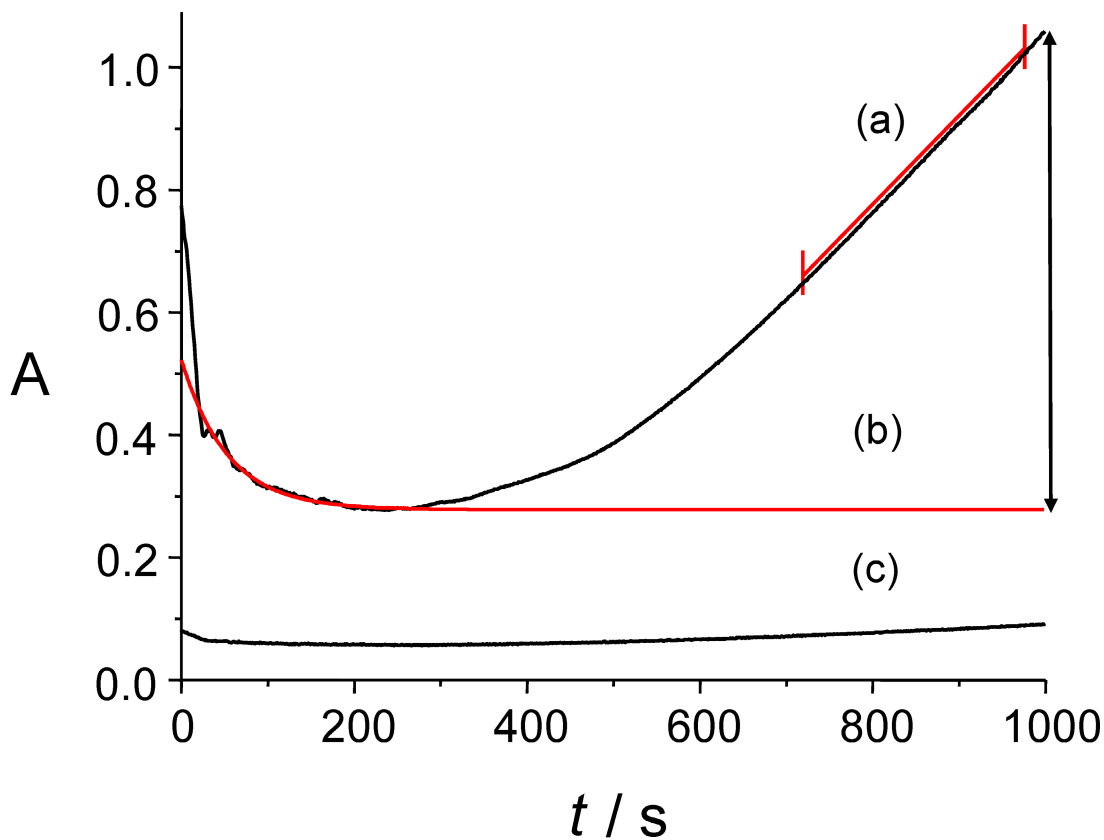

**Figure 4**. Examples of the output signal evolution for two different choices of the three varied inputs, for the optimized system, and an illustration of the exponential fitting to subtract the starch background. The absorbance, A, is plotted as a function of time. (a) The maximal-input combination, $(x_1, x_2, x_3) = (1,1,1)$. The red segment represents a typical selection of a linear part for the slope signal calculation. (b) Background fitting for the $(x_1, x_2, x_3) = (1,1,1)$ curve. The vertical double-arrow represents the net signal, denoted $[\Delta A_{NADH}(t_{max})]_{max}$ in the text, at time $t_{max} = 1000$ s. (c) A typical response for the input combination $(x_1, x_2, x_3) = (0.05, 1, 1)$. As detailed in Sec. 4.2, for the optimized set this was the measurement which was the closest to the Boolean-logic values $(x_1, x_2, x_3) = (0,1,1)$ for the varied inputs.

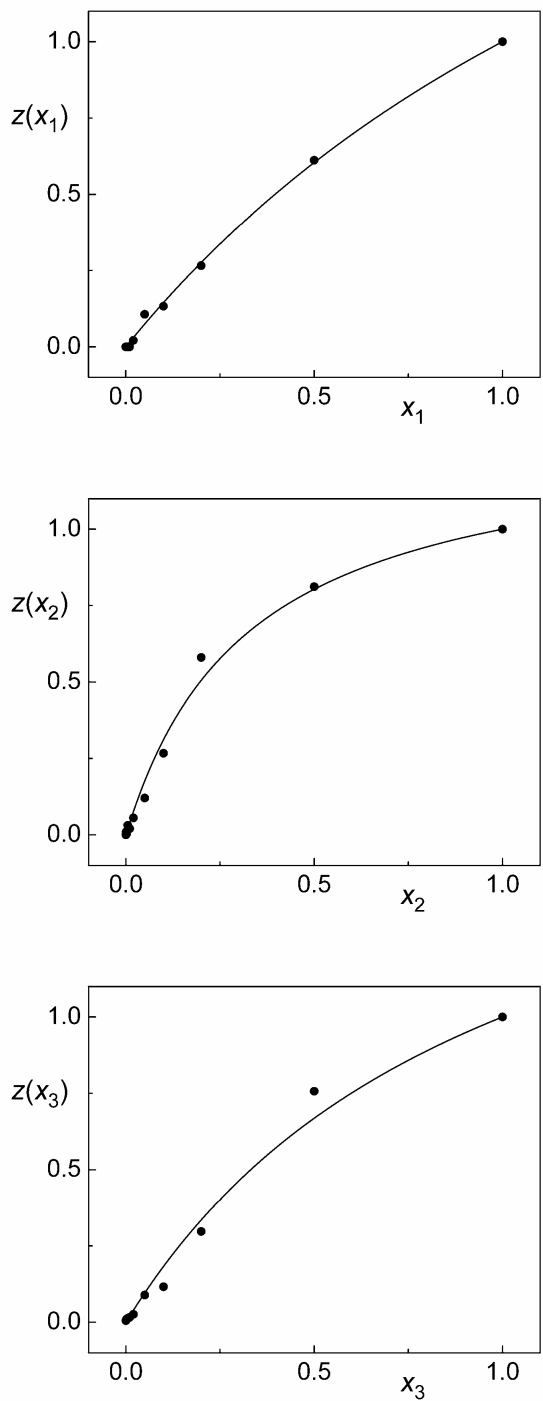

**Figure 5**. Experimental data for the initially selected parameter set, with the signal defined via the slope of the absorbance. The data for inputs *x* varied at gates 1, 2, 3, are shown, respectively, from top to bottom. The data were fitted to the single-parameter curves, Eq. 16-18, with the results presented in Table 1.

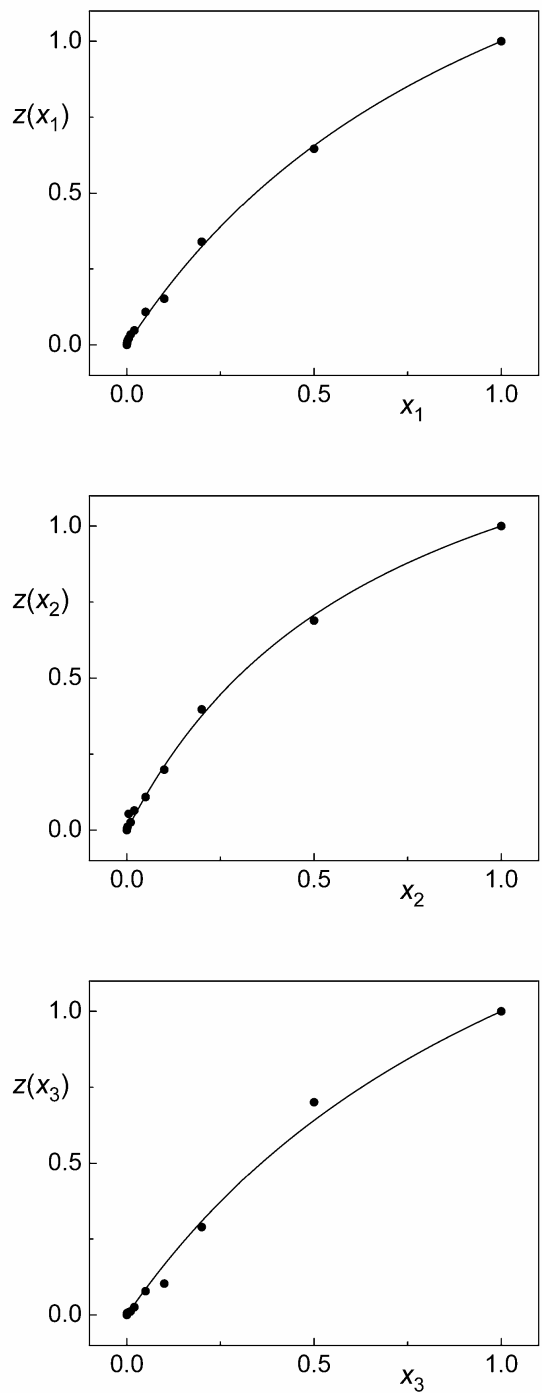

**Figure 6**. Experimental data for the initially selected parameter set, with the signal defined via the fixed-time values the absorbance. The data for inputs *x* varied at gates 1, 2, 3, are shown, respectively, from top to bottom. The data were fitted to the single-parameter curves, Eq. 16-18, with the results presented in Table 1.

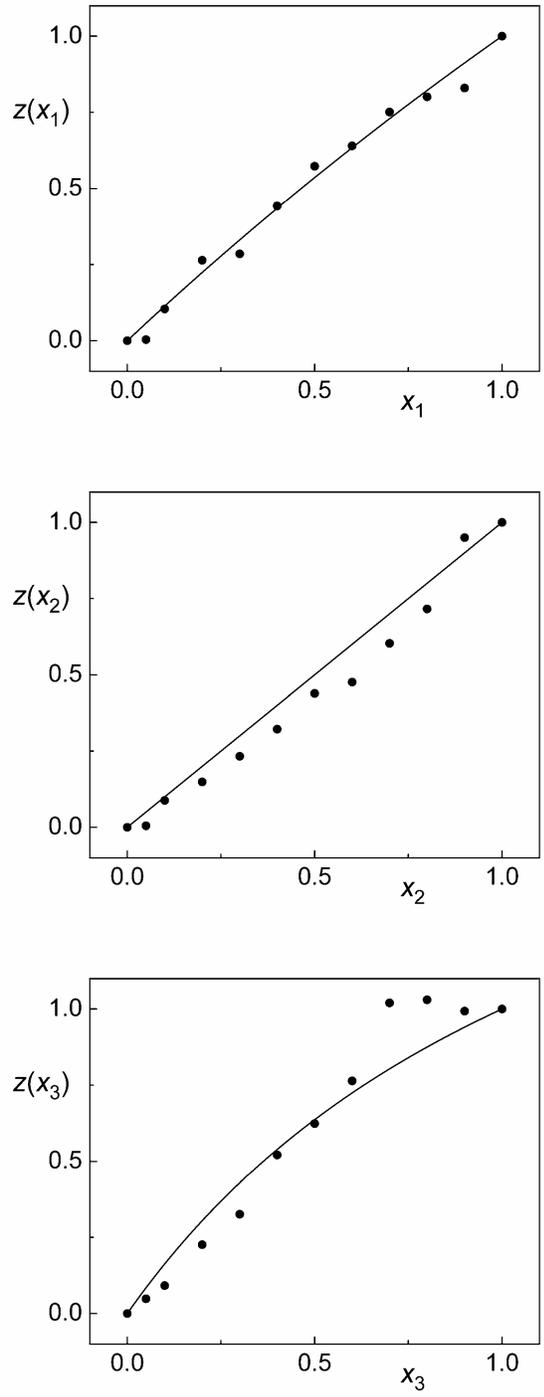

**Figure 7**. Experimental data for the modified parameter set, with the signal defined via the slope of the absorbance. The data for inputs $x$ varied at gates 1, 2, 3, are shown, respectively, from top to bottom. The data for $z(x_{1,3})$ were fitted to the single-parameter curves, Eq. 16, 18, with the results presented in Table 3. With the constraint of the selected fitting function, Eq. 17, the data for $z(x_2)$ are best approximated by the straight line connecting the two logic points, as shown, obtained for the largest possible value, 1, of the fit parameter, from the allowed range $(0,1]$, see Eq. 14.

**Figure 8**. Experimental data for the modified parameter set, with the signal defined via the fixed-time values the absorbance. The data for inputs $x$ varied at gates 1, 2, 3, are shown, respectively, from top to bottom. The data for $z(x_3)$ were fitted to the single-parameter curve, Eq. 18, with the result presented in Table 3. With the constraint of the selected fitting function, Eq. 16-17, the data for $z(x_{1,2})$ are best approximated by the straight lines connecting the two logic points, as shown, obtained for the largest possible value, 1, of the respective fit parameters, from the allowed range (0,1], see Eq. 14.

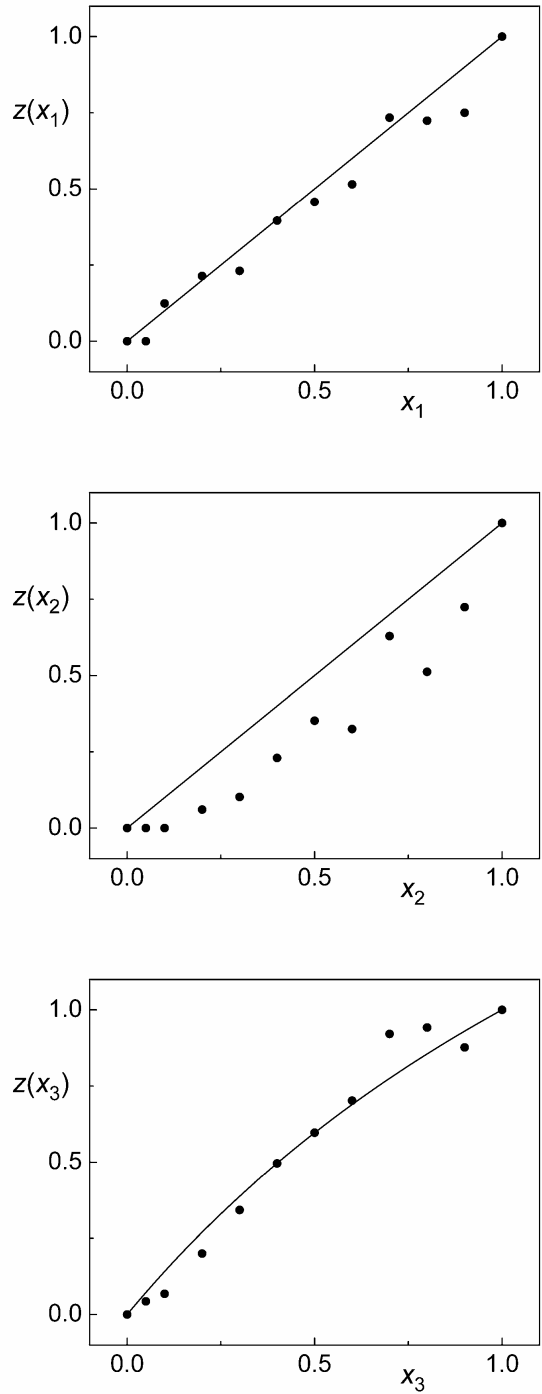

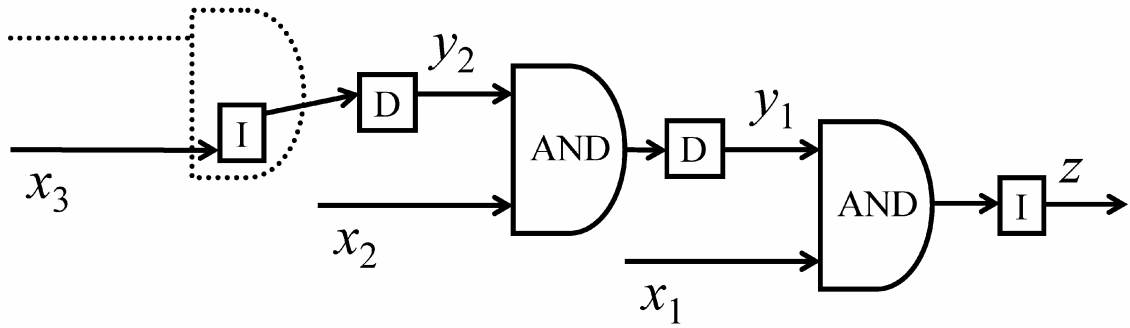

**Figure 9**. An alternative representation of the logic network, with the use of the identity function, I, and identity with time-delay, denoted by D.